\documentclass[12pt]{article}

\usepackage{amssymb}
\usepackage{amsmath}
\usepackage{amscd}
\usepackage{latexsym}
\usepackage{graphicx}

\usepackage{cite}

\newcommand{\be}{\begin{equation}}
\newcommand{\ee}{\end{equation}}

\newcommand{\prt}{\partial}

\newcommand{\oa}{\overline a}
\newcommand{\ob}{\overline b}
\newcommand{\ra}{\rightarrow}
\newcommand{\bt}{\beta}

\newcommand{\ep}{\varepsilon}
\newcommand{\al}{\alpha}

\newcommand{\gm}{\gamma}

\begin{document}

\begin{center}

{\Large{\bf Self-similarly corrected Pad\'{e} approximants for nonlinear equations} \\ [5mm]

S. Gluzman$^1$ and V.I. Yukalov$^{2,3}$ } \\ [3mm]

$^1$Bathurst St. 3000, 606, Ontario M6B 3B4, Toronto, Canada \\
E-mail: simongluzmannew@gmail.com  \\ [2mm]

$^2${\it Bogolubov Laboratory of Theoretical Physics, \\
Joint Institute for Nuclear Research, Dubna 141980, Russia \\ [2mm]

$^3$Instituto de Fisica de S\~ao Carlos, Universidade de S\~ao Paulo, \\
CP 369,  S\~ao Carlos 13560-970, S\~ao Paulo, Brazil  } \\ 
E-mail: yukalov@theor.jinr.ru \\ [5mm]

\end{center}

\vskip 2cm

\begin{abstract}
We consider the problem of finding approximate analytical solutions for nonlinear
equations typical of physics applications. The emphasis is on the modification of the 
method of Pad\'e approximants that are known to provide the best approximation for 
the class of rational functions, but do not provide sufficient accuracy or cannot 
be applied at all for those nonlinear problems, whose solutions exhibit behaviour 
characterized by irrational functions. In order to improve the accuracy, we suggest 
a method of self-similarly corrected Pad\'e approximants, taking into account 
irrational functional behaviour. The idea of the method is in representing the sought 
solution as a product of two factors, one of which is given by a self-similar root 
approximant, responsible for irrational functional behaviour, and the other being 
a Pad\'e approximant corresponding to a rational function. The efficiency of the method
is illustrated by constructing very accurate solutions for nonlinear differential 
equations. A thorough investigation is given proving that the suggested method is more 
accurate than the method of standard Pad\'e approximants.   
\end{abstract}



\newpage

\section{Introduction}

Nonlinear problems are widespread in different branches of physics. As a rule, such problems 
can very rarely be solved exactly. But it is often useful to have in hands an approximate 
solution in analytic form. Various ways of constructing asymptotic approximate solutions
for nonlinear equations have recently been reviewed by He \cite{He_1}. Another way of 
constructing approximate solutions is by using the method of Pad\'{e} approximants 
\cite{Baker_1}. It is well known that the latter provide the best approximation for the class 
of rational functions. However, approximating irrational functions by Pad\'{e} approximants 
can result in pore accuracy and sometimes is not possible at all 
(see, e.g., Refs. \cite{Epele_3,Gluzman_7}). When solutions to nonlinear equations exhibit 
irrational functional behavior, to reach a good accuracy, it is necessary to go to high-order 
Pad\'{e} approximants. In these cases Pad\'{e} approximants may display unphysical poles and 
zeroes \cite{Baker_1,Gluzman_7,Gluzman_8}. 

In the present paper we suggest a method for overcoming the above mentioned problems, when 
approximating irrational functions. The idea of the method is to correct Pad\'e approximants 
by combining them with the other type of approximants that would take into account the 
irrational behavior of the sought functions. We show that the method is applicable not only 
to rather simple problems, but can be used for solving nonlinear differential equations. 
By a careful comparison, we prove that the corrected Pad\'{e} approximants are essentially 
more accurate than the standard Pad\'{e} approximants.  

The outline of the paper is as follows. In sec. 2, we present the main ideas of the suggested 
method and explain why self-similar root approximants can efficiently describe the behaviour 
of irrational functions. In sec. 3, we solve the Thomas-Fermi equation, then in sec. 4, the 
nonlinear Schr\"odinger equation, and in sec. 5, the Ruina-Dieterich equation. In all the 
cases, we thoroughly compare our solutions with the standard Pad\'{e} approximants and show 
that the corrected Pad\'{e} approximants are more accurate. In Sec. 6, we discuss possible 
extensions of the method. One such an extension is the procedure of finding the 
large-variable behavior of a function from its small-variable asymptotic form. We also 
discuss the possible generalization of the method for partial differential equations. 
Section 7 concludes.

\section{Self-similarly corrected Pad\'{e} approximants}

The straightforward way of how it would be possible to improve the accuracy of approximants
for irrational solutions is to take into account this irrational behavior by an appropriate
approximation procedure. Approximate solutions in the class of irrational functions can be
constructed by invoking self-similar approximation theory that was advanced in Refs. 
\cite{Yukalov_8,Yukalov_11,Yukalov_12,Yukalov_13,Yukalov_14,Yukalov_15}. This approach 
combines the ideas of dynamical theory, optimal control theory, and renormalization group 
\cite{Struble_15,Irwin_16,Farmer_17,Peitgen_18,Yukalov_19}. A convenient type of approximants, 
resulting from the self-similar approximation theory and representing well irrational 
functional forms, is given by self-similar root approximants
\cite{Yukalov_20,Gluzman_21,Yukalov_22,Gluzman_23,Yukalov_24,Gluzman_9}. In principle, it 
could be possible to employ the self-similar approximants as such. But it would, probably, 
be a pity to forget the well developed techniques of Pad\'{e} approximants. So, the important 
question is whether it would be possible to modify Pad\'{e} approximants in such a way that 
to improve their accuracy for the class of irrational functions.

The desired modification could be done by splitting the sought solution into two factors, one, 
represented by root approximants or factor approximants, taking care of the irrational part of 
the solution, and the other being a Pad\'{e} approximant, characterizing the rational part of 
the solution. The so corrected Pad\'{e} approximants would be applicable to a larger class of 
problems and be well defined even for those cases, where the standard Pad\'{e} approximants 
could not be used. Below, we present the main ideas of the method. 

Suppose we wish to find a real function $f(x)$ of a real variable $x$, which is defined
by a complicated nonlinear equation, whose exact solution is not available. But it is
often possible to derive an asymptotic form of the sought solution, say, at small values
of the variable,
\be
\label{1}
 f(x) \simeq f_k(x) \qquad ( x \ra 0 ) \;  ,
\ee
where it is presented by an expansion in powers of $x$:
\be
\label{2}
f_k(x) = f_0(x) \left ( 1 + \sum_{n=1}^k a_n x^n \right ) \;   ,
\ee
with $f_0(x)$ being a known function.

And let the asymptotic form of the sought function at large values of the variable,
\be
\label{3}
f(x) \simeq f^{(p)}(x) \qquad ( x \ra \infty ) \;   ,
\ee
be given as an expansion
\be
\label{4}
 f^{(p)}(x) = \sum_{n=1}^p b_n x^{\bt_n}  ,
\ee
with the descending order of the powers,
$$
 \bt_{n+1} < \bt_n \qquad ( n = 1,2,\ldots,p-1) \;  .
$$

Assume that we can construct an irrational function $f_{irr}^*(x)$ satisfying the
boundary conditions represented by the above asymptotic expansions,
$$
f_{irr}^*(x) \simeq f_k(x) \qquad ( x \ra 0 ) \; ,
$$
\be
\label{5}
f_{irr}^*(x) \simeq   f^{(p)}(x) \qquad ( x \ra \infty ) \;  .
\ee

Then we introduce a correcting function
\be
\label{6}
C_k(x) \equiv \frac{f_k(x)}{f_{irr}^*(x)}
\ee
that, being expanded in powers of $x$, acquires the form
\be
\label{7}
 C_k(x) \simeq 1 + \sum_{n=1}^k d_n x^n \qquad ( x \ra 0 ) \;  .
\ee
The first term here is one, since the function $f_{irr}^*(x)$ tends to $f_k(x)$ at small $x$.

On the basis of expansion (\ref{7}), it is straightforward to generate a diagonal
Pad\'{e} approximant $P_{N/N}(x)$, such that
$$
 P_{N/N}(x) \simeq C_k(x)  \qquad ( x \ra 0 ) \;  .
$$
Then the approximate solution is represented as the product
\be
\label{8}
 f_k^*(x) = f_{irr}^*(x) P_{N/N}(x) \;  .
\ee

If the irrational factor $f_{irr}^*(x)$ is assumed to satisfy the boundary conditions (\ref{5}),
then the Pad\'{e} approximant has to obey the boundary conditions
\be
\label{9}
 P_{N/N}(0) = P_{N/N}(\infty) = 1 \; .
\ee
However, in general, it is not compulsory that conditions (\ref{5}) and (\ref{9}) be valid
separately, but is sufficient that the product solution (\ref{8}) as a whole would satisfy
the asymptotic conditions
$$
f_k^*(x) \simeq f_k(x) \qquad ( x \ra 0 ) \; ,
$$
\be
\label{10}
 f_k^*(x) \simeq f^{(p)}(x)  \qquad ( x \ra \infty ) \;  .
\ee

The so constructed approximate solution (\ref{8}) takes into account the irrational
functional behaviour of the sought solution and allows one to invoke the techniques
of Pad\'{e} approximants for reaching good accuracy of this solution.

The irrational factor can be presented in different forms. Here we show that a very
convenient form is provided by self-similar root approximants
\cite{Yukalov_20,Gluzman_21,Yukalov_22,Gluzman_23,Yukalov_24,Gluzman_9} resulting from the 
self-similar approximation theory
\cite{Yukalov_8,Yukalov_11,Yukalov_12,Yukalov_13,Yukalov_14,Yukalov_15}.
 
Suppose we have the small-variable expansion (\ref{2}) that can be divergent for finite
values of the variable $x$. To make sense of divergent series, it is necessary to introduce
control parameters $s_k$ generating control functions $s_k(x)$ such that the series of the
terms $f_k(x)$ is renormalized into the series of the expressions $F_k(x,s_k(x))$ that become
convergent. This renormalization can be formalized as
\be
\label{S0}
 F_k(x,s_k) = \hat R[s_k]f_k(x) \;  .
\ee
The control functions should be defined from asymptotic and optimization conditions.

The second idea is the reformulation of the study of the sequence $\{F_k\}$ into the language
of dynamical theory in order to resort to the powerful techniques of the latter. For this
purpose, it is necessary to introduce a sequence of endomorphisms $y_k$ in the space of 
approximations, which is bijective to the approximation sequence $\{F_k\}$. These endomorphisms 
are constructed in the following way. We define the expansion function $x = x_k(f)$ by the 
reonomic constraint
\be
\label{S1}
 F_0(x,s_k(x)) = f \; , \qquad x = x_k(f) \;  .
\ee
This makes it possible to introduce the endomorphism
\be
\label{S2}
 y_k(f) \equiv F_k(x_k(f),s_k( x_k(f) )) 
\ee
acting in the space of approximations. Treating the passage from an approximation $y_k$ 
to another approximation $y_{k+p}$ as the motion with respect to the approximation order $k$, 
we can consider the sequence of endomorphisms $\{y_k\}$ in the space of approximations as 
the trajectory of a dynamical system, with $k$ playing the role of discrete time.

Since by construction the sequences $\{F_k\}$ and $\{y_k\}$ are bijective, the
convergence of $\{F_k\}$ implies the existence of an attractive fixed point $y^*$
for the sequence $\{y_k\}$, where $y_k(y^*) = y^*$. In the vicinity of a fixed point,
the dynamical system enjoys the property of self-similarity, such that the relation
is valid:
\be
\label{S3}
 y_{k+p}(f) = y_k(y_p(f)) \;  .
\ee
This relation defines a cascade, that is a discrete dynamical system.

The next step is the embedding of the cascade into a flow, whose trajectory $\{y(t,f)\}$
passes through all points of the cascade trajectory,
\be
\label{S4}
 y(t,f) = y_k(f) \qquad ( k = t ) \;  ,
\ee
so that the self-similar relation (\ref{S3}) be preserved,
\be
\label{S5}
 y(t+t',f) = y(t,y(t',f) ) \;  .
\ee

The latter relation can be rewritten in the differential form of the Lie equation
\be
\label{S6}
 \frac{\prt}{\prt t} \; y(t,f) = v(y(t,f)) \;  ,
\ee
where $v$ is the velocity of the dynamical system. Integrating the Lie equation
between $y_k$ and $y^*_k$, with the cascade velocity
\be
\label{S7}
 v_k(f) = y_k(f) - y_{k-1}(f) \;  ,
\ee
defines an approximate fixed point $y^*_k$. Performing the transformation inverse
to  (\ref{S0}), we obtain the self-similar approximant
\be
\label{S8}
  f_k^*(x) = \hat R^{-1}[s_k] y_k^*(F_0(x,s_k(x) )) \; .
\ee

For deriving self-similar root approximants, we introduce control functions through
the fractal transform
\be
\label{S9}
F_k(x,s_k) = x^{s_k} f_k(x) \;   .
\ee
This transform allows us to extract the scaling properties of the resulting sequence of
approximants.

Then we integrate the Lie equation (\ref{S6}) between the quasi-fixed points
$y^*_{k-1}$ and $y^*_k$, which gives the recurrent relation
\be
\label{S10}
 y_k^* = y_k^*(y_{k-1}^*) \;  .
\ee
Iterating this recurrent relation $k-1$ time yields
\be
\label{S11}
 y_k^* = y_k^*\left( y_{k-1}^* \left( y_{k-2}^* \ldots \left( y_0^*
\right) \right) \right) \;  .
\ee
With the inverse fractal transform
\be
\label{S12}
 f_k^*(x) = x^{-s_k} y^*_k \left( x^{s_k} \right) \;  ,
\ee
we come to the recurrent relation
\be
\label{S13}
 f_k^* = f_k^*\left( f_{k-1}^* \left( f_{k-2}^* \ldots \left( f_0^*
\right) \right) \right) \;  .
\ee

Accomplishing the described steps for the asymptotic expansion of the sought function
in equation (\ref{2}), we obtain the self-similar root approximant
\be
\label{R1}
 f_k^*(x) = f_0(x) \left ( \left ( (1 + A_1 x)^{n_1} + A_2 x^2 \right )^{n_2} +
\ldots + A_k x^k \right )^{n_k} \; .
\ee
The parameters $A_i$ and powers $n_i$ are defined so that to satisfy the boundary
asymptotic conditions (\ref{10}). It has been shown \cite{Yukalov_25} that this procedure 
can be uniquely defined. Thus, if $f_0(x) = A x^\alpha$ and $p < k$, then the boundary 
conditions uniquely define all parameters of the root approximant (\ref{R1}), prescribing 
the powers by the equalities
$$
n_j = \frac{j+1}{j} \qquad ( j = 1,2,\ldots,k-p) \; ,
$$
$$
jn_j = j + 1 + \bt_{k-j+1} - \bt_{k-j} \qquad ( j = k-p+1, k-p+2, \ldots , k-1 ) \; ,
$$
$$
k n_k = \bt_1 - \al \; .
$$
Of course, the order of the root approximant is to be lower than the order of expansion
(\ref{2}), since we need the remaining terms for constructing Pad\'{e} approximants.
As is evident, the root approximant, having irrational form, can efficiently take into
account the corresponding features of the sought function.

One can remember that some irrational functions that yield the asymptotic expansions in
noninteger powers,
$$
 f_k(x) = f_0(x) \left ( 1 + \sum_{n=1}^k c_n x^{\gm n} \right ) \;  ,
$$
where $\gamma$ is not integer, can easily be approximated by the Pad\'{e} approximants
of the form
$$
P_{M/N}(x) = \frac{\sum_{m=0}^M a_m x^{\gm m} }{ \sum_{n=0}^N b_n x^{\gm n} } \; .
$$
As is clear, this kind of approximants is nothing but the standard Pad\'{e} approximants with
a straightforward change of the variable $z = x^\gamma$. Hence the above approximant becomes 
rational with respect to the variable $z$. Respectively, this approximant represents a
function that is rational with respect to $z$. This type of irrationality that can be easily 
reduced to a rational form can be called the reducible irrationality. This fact is well known 
and we use it without special comments.

Contrary to this, the root approximant (\ref{R1}) cannot be reduced to a rational
form by a simple change of the variable. In that sense, it describes a more general,
nonreducible irrationality. While the correcting function (\ref{7}) can have an
expansion in noninteger powers, which will lead to the related Pad\'{e} approximant
composed of polynomials of noninteger powers.

Taking into account the irrational behavior by means of root approximants accelerates the
convergence of the approximation procedure. As is mentioned above, in general, Pad\'{e}
approximants can also approximate irrational functions. At lower orders, such Pad\'{e}
approximants oscillate, or give insufficiently accurate results, or display unphysical poles
and zeros. Sometimes (although not always), as stated earlier, these drawbacks can 
simultaneously disappear by resorting to higher-order Pad\'{e} approximants. In fact, the 
reader should be reminded that the only acceptable results from the table of Pad\'{e} 
approximants are those that have converged/stabilized, and these generally necessitate 
resorting to high-orders. However, taking into account the irrational behavior from the very 
beginning allows a quicker attainment of the desired accuracy (i.e. with lower orders).

\section{Thomas-Fermi equation}

The Thomas-Fermi equation describes the screened Coulomb potential caused by
a heavy charged nucleus surrounded by a cloud of electrons \cite{Spruch_26}. The
equation reads as
\be
\label{11}
  \frac{d^2 f(x)}{dx^2} = \frac{f^{3/2}(x)}{\sqrt{x}} \; .
\ee
For neutral atoms, the boundary conditions are
\be
\label{12}
 f(0) = 1 \; , \qquad f(\infty) = 0 \;  .
\ee
At small $x \ra 0$, the solution to the equation can be written \cite{Fermi_27,Baker_28}
in the form of the expansion
$$
 f(x) \simeq 1 - Bx \; + \; \frac{1}{3}\; x^3 \; - \;\frac{2B}{15}\; x^4 \; + 
$$
$$
+ \;
x^{3/2}
\left [ \frac{4}{3} \; - \; \frac{2B}{5} \; x \; + \; \frac{3B^2}{70}\; x^2 \; + \;
\left ( \frac{2}{27} \; + \;\frac{B^3}{252} \right ) x^3 \right ] \; ,
$$
where $B = 1.588071$ is found numerically \cite{Pindov_29}. This gives
\be
\label{13}
 f(x) \simeq 1 + \sum_{n=1}^9 a_n x^{n/2} \qquad ( x \ra 0 ) \;  ,
\ee
with the coefficients
$$
a_1 = 0 \; , \quad a_2 = -1.588071 \; , \quad a_3 = \frac{4}{3} \; , \quad a_4 = 0 \; ,
\quad a_5 = - 0.635228 \; ,
$$
$$
a_6 = \frac{1}{3} \; , \qquad a_7 = 0.108084 \; , \qquad
 a_8 = - 0.211743 \; , \qquad a_9 = 0.0899672 \;  .
$$

At large $x$, the asymptotic behaviour is \cite{Bender_30}
\be
\label{14}
 f(x) \simeq b_1 x^\bt_1 + b_2 x^{\bt_2} \qquad ( x \ra \infty ) \;  ,
\ee
where
$$
 b_1 = 144 \; , \qquad \bt_1 = -3 \; , \qquad b_2 = 1911.02 \; , \qquad
\bt_2 = -3.772 \;  .
$$

Following the general method of constructing self-similar root approximants, using expansion 
(\ref{13}), to second order, we have
\be
\label{15}
 f_2^*(x) = \left ( ( 1 + A_1 x)^{n_1} + A_2 x^{3/2} \right )^{-2} \;  ,
\ee
with the parameters defined by the asymptotic boundary conditions (\ref{10}),
$$
 A_1 = 0.443153 \; , \qquad A_2 = 0.0833333 \; , \qquad n_1 = 0.727998 \;  .
$$
To third order, we find
\be
\label{16}
 f_3^*(x) = \left ( \left ( ( 1 + B_1 x)^{n_1} + B_2 x^{3/2} \right )^{n_2} + B_3 x^2
\right )^{-3/2} \;  ,
\ee
where the parameters are
$$
B_1 = 1.7764 \; , \qquad n_1 = 0.727998 \; ,
$$
$$
B_2 = 0.250555 \; , \qquad
n_2 = 0.818665 \; , \qquad B_3 = 0.0363992 \; .
$$
As the irrational factor, we can take
\be
\label{17}
  f_{irr}^*(x) = \frac{1}{2} \left [ f_2^*(x) + f_3^*(x) \right ] \; .
\ee

Introducing the correcting function
\be
\label{18}
 C_8(x) = \frac{f_8(x)}{f_{irr}^*(x)} \;  ,
\ee
which is defined by the eight order of expansion (\ref{13}), we obtain the
small-variable expansion
\be
\label{19}
 C_8(x) \simeq 1 + \sum_{n=1}^8 d_n x^{n/2} \qquad (x \ra 0) \;  ,
\ee
with the coefficients
$$
d_1 = 0 \; , \qquad d_2 = -0.471421 \; , \qquad d_3 = 1.57051 \; , \qquad
d_4 = - 2.01043 \; ,
$$
$$
d_5 = 0.482756 \; , \quad d_6 = 1.41347 \; , \quad d_7 = 0.838164 \; ,
\quad d_8 = -1.07168 \;  .
$$

On the basis of expansion (\ref{19}), we define the Pad\'{e} approximant
\be
\label{20}
P_{4/4}(x) = \frac{1 + \oa_1 x^{1/2} + \oa_2 x +\oa_3 x^{3/2} + \oa_4 x^2}
{1 + \ob_1 x^{1/2} + \ob_2 x +\ob_3 x^{3/2} + \ob_4 x^2}
\ee
satisfying the boundary conditions
\be
\label{21}
 P_{4/4}(0) =  P_{4/4}(\infty) = 1 \; .
\ee
The parameters of the approximant (\ref{20}) are
$$
\oa_1 = 2.79159 \; , \qquad \oa_2 = 4.56393 \; , \qquad \oa_3 = 6.14842 \; , \qquad
\oa_4 = 4.01834 \; ,
$$
$$
 \ob_1 = 2.79159 \; , \qquad \ob_2 = 5.03535 \; , \qquad \ob_3 = 5.89392 \; , \qquad
\ob_4 = 4.01834 \;  .
$$

The sought corrected Pad\'{e} approximant becomes
\be
\label{22}
 f_8^*(x) = f_{irr}^*(x) P_{4/4}(x) \;  .
\ee
The accuracy of the approximant can be characterized by the relative errors
\be
\label{23}
 \ep_8^*(x) \equiv \frac{f_8^*(x) - f(x)}{f(x)} \times 100 \% \;  ,
\ee
where $f(x)$ is the exact numerical solution of the Thomas-Fermi equation (\ref{11}).

For completeness, we also present the lower-order corrected approximants, $f_4^*(x)$,
\be
\label{2222}
 f_4^*(x) = f_{irr}^*(x) P_{2/2}(x) \;  .
\ee
with
$$
P_{2/2}(x)=\frac{1+3.33143 \sqrt{x}+6.36239 x}{1+3.33143 \sqrt{x}+6.83382 x} \; ,
$$
and $f_6^*(x)$,
\be
\label{222}
 f_6^*(x) = f_{irr}^*(x) P_{3/3}(x) \; ,
\ee
with
$$
P_{3/3}(x)=\frac{1+2.37227 \sqrt{x}+3.16702 x+3.48058 x^{3/2}}
{1+2.37227 \sqrt{x}+3.63844 x+3.02841 x^{3/2}} \; .
$$
The related values of the approximants and their errors are shown in Table 1.

Among empirical analytic forms, it is worth mentioning the empirical Sommerfeld
solution \cite{Sommerfeld_31}
\be
\label{24}
f_S(x) = \frac{1}{(1+0.278343 x^{0.772002})^{3.886} }
\ee
and the Andrianov-Awrejcewicz empirical solution \cite{Andrianov_32},
$$
f_A(x) =\left [ 1 + 0.1336 x^{1/2} - 1.3038 x + 0.9598 x^{3/2} -
0.2523 x^2 + x^{5/2} \right ] \times
$$
\be
\label{25}
 \times
\left [ 1 + 0.1336 x^{1/2} + 0.2842 x - 0.1614 x^{3/2} +
0.0209 x^2 + F(x) x^{5/2} \right ]^{-1}   \;  ,
\ee
in which
$$
F = \left [ 1 + \frac{0.2783x}{(1+x)^{0.228} } \right ]^{3.886} \; .
$$
The relative errors $\varepsilon_S(x)$ and $\varepsilon_A(x)$ can be defined similarly
to the relative errors (\ref{23}).

In order to compare our results with the standard Pad\'{e} approximants, defined so that 
to satisfy the boundary conditions (\ref{10}), we calculate the Pad\'{e} approximants having 
no unphysical poles. To construct $P_{1/4}(x)$ and $P_{2/5}(x)$, we add to the series an 
additional zero term. Thus we obtain
$$
P_{0/3}(x)=\frac{Q_0(x)}{Q_3(x)} \; ,
$$
where
$$
Q_0(x) = 1 \; , 
$$
$$
Q_3(x) = 1 + x (5.44951 x^2 - 3.59963 x^{3/2} + 2.52197 x - 1.33333 \sqrt{x} + 1.58807 ) \; ,
$$

$$
P_{1/4}(x)= \frac{Q_1(x)}{Q_4(x)} \; ,
$$
where
$$
Q_1(x) = 1 + 7.29513 \sqrt{x} + 8.70365 x \; ,
$$
$$
Q_4(x) = 
1+7.29513 \sqrt{x} + 10.2917 x + 10.2519 x^{3/2} + 6.61714 x^2 + 
$$
$$
+ 3.19361 x^{5/2} + 1.14009 x^3 + 0.246624 x^{7/2} + 0.0573431 x^4 \; ,
$$
$$
P_{2/5}(x)= \frac{Q_2(x)}{Q_5(x)} \; ,
$$
where
$$
Q_2(x) = 1 + 0.0611225 \sqrt{x} - 0.75871 x + 2.75597 x^{3/2} + 0.957022 x^2 \; ,
$$
$$
Q_5(x) = 
1 + 0.0611225 \sqrt{x} + 0.829361 x + 1.51971 x^{3/2} + 2.19261 x^2 + 
$$
$$
+ 1.94282 x^{5/2} + 1.16124 x^3 + 0.560225 x^{7/2} +
$$
$$
+ 0.147749 x^4 + 0.0609348 x^{9/2} - 0.00321342 x^5 \; .
$$

We have also constructed $P_{3/6}(x)$, but its accuracy is rather bad, with the maximal error 
around $80\%$ at $x=1000$. 

The other admissible sequence of approximants is given by
$$
 P_{1/7}(x) = \frac{1+a_1x^{1/2}}{1+\sum_{n=1}^7 b_nx^{n/2}} \;  ,
$$
with the parameters
$$
 a_1 = 1.502670=b_1 \; , \quad b_2 = 1.588071 \; , \quad b_3 = 1.053015 \; ,
\quad b_4 = 0.518408 \; ,
$$
$$
b_5 = 0.190063 \; , \qquad b_6 = 0.040455 \; , \qquad b_7 = 0.010435 \; ,
$$
and the Pad\'{e} approximant
$$
  P_{2/8}(x) = \frac{1+a_1x^{1/2}+a_2x}{1+\sum_{n=1}^8 b_nx^{n/2}} \;  ,
$$
with the parameters
$$
a_1 = -8.448419=b_1 \; , \qquad a_2 = -14.953104 \; , \qquad b_2 = -13.365089 \; ,
$$
$$
b_3 = -14.750023 \; , \qquad b_4 = -9.960151 \; , \quad  b_5 = -4.968737 \; ,
$$
$$
b_6 = -1.850739 \; , \qquad b_7 = -0.392334 \; ,  \qquad b_8 = -0.103841 \;  .
$$

The results for the Sommerfeld solution $f_S(x)$, the Andrianov-Awrejscewicz solution $f_A(x)$,
and for the Pad\'{e} approximants $P_{0/3}(x)$ and $P_{1/7}(x)$, with their errors, are shown
in Table 2. The approximant $P_{2/8}(x)$ practically coincides with $P_{1/7}(x)$, because of
which it is not shown. The percentage errors of the Pad\'{e} approximants $P_{M/N}(x)$ are
denoted as
$$
 \ep_{M/N}(x) \equiv \frac{P_{M/N}(x) - f(x)}{f(x)} \times 100\% \;  .
$$

There have been suggested approximate solutions derived by employing the Lagrange variational
techniques \cite{Desaix_32,Bougoffa_33,Sierra_34}. However, these solutions are not sufficiently
accurate, yielding relative errors of order $100 \%$. The empirical  Andrianov-Awrejcewicz
solution \cite{Andrianov_32} has been the most accurate of analytical solutions known till now.
But our solution (\ref{22}) is an order more accurate.

\begin{table}
\caption{Percentage errors of the self-similarly corrected Pad\'e approximants
$f_4^*(x)$, $f_6^*(x)$, and $f_8^*(x)$.}
\label{Table 1}       
\centering
\begin{tabular}{lllllll}
\hline\noalign{\smallskip}
$x$ & $f_4^*(x)$ & $\ep_4^*(x)$ & $f_6^*(x)$ & $\ep_6^*(x)$ & $f_8^*(x)$ & $\ep_8^*(x)$  \\ 
\noalign{\smallskip}\hline\noalign{\smallskip} 
0.1 & 0.880            & $-$0.23 & 0.882    & 0.012 & 0.882      &  0.0019  \\
1   & 0.411            & $-$3.06 & 0.428    & 1.018 & 0.424      &  0.04       \\
40  & 0.00103          & $-$7.49 & 0.00121  & 8.99  & 0.00111    & $-$0.63    \\
100 & 0.0000933        & $-$6.89 & 0.000112 & 11.46 & 0.000100   &  0.095   \\
1000& 1.3$\times10^{-7}$ & $-$6.80 & 1.5$\times10^{-7}$ & 13.88 & 1.4$\times10^{-7}$ &  0.16  \\ 
\noalign{\smallskip}\hline
\end{tabular} 
\end{table}

\begin{table}
\caption{Sommerfeld solution $f_S(x)$, the Andrianov-Awrejscewicz solution $f_A(x)$,
and the Pad\'e approximant $P_{1/7}(x)$. The approximant $P_{2/8}(x)$ practically coincides 
with $P_{1/7}(x)$.}
\label{Table 2}      
\centering
\begin{tabular}{lllllllllll}
\hline\noalign{\smallskip}
$x$ &0.1&1&40&100&1000  \\
\noalign{\smallskip}\hline\noalign{\smallskip}
$P_{0/3}(x)$ &0.880&0.178&3.2$\times10^{-6}$& 2$\times10^{-7}$&1.9$\times10^{-10}$\\
$\ep_{0/3}(x)$&-0.15& $-$58.083 &-$99.72$&99.80&-99.86\\
$f_S(x)$&0.836&0.385&0.00108&0.0000985&1.3$\times10^{-7}$\\
$\ep_S(x)$ &-5.14&-9.18&-3.17&-1.7&-0.33\\
$f_A(x)$  &0.882&0.431&0.00108& 0.0000977&1.3$\times10^{-7}$\\
$\ep_A(x)$ &0.076&1.64&-3.41&-2.53&-0.88 \\
$P_{1/7}(x)$  &0.882&0.424&0.00106&0.0000942&1.3$\times10^{-7}$\\
$\ep_{1/7}(x)$ &$-$0.00034 &$-$0.01&-4.82&-6.04&-4.76\\
\noalign{\smallskip}\hline
\end{tabular}
\end{table}

The Pad\'e approximants $P_{0/3}(x)$, $P_{1/4}(x)$, and $P_{2/5}(x)$ are shown in Table 3.
The approximant $P_{1/4}(x)$ is of the same quality as the Andrianov-Awrejscewicz solution.

\begin{table}
\caption{The Pad\'e approximants $P_{0/3}(x)$, $P_{1/4}(x)$, $P_{2/5}(x)$. Numerical solution 
from \cite{Bender_30}, named as ``exact'', is shown as well.
}
\label{Table 3}      
\centering
\begin{tabular}{lllllll}
\hline\noalign{\smallskip}
$x$ &0.1&1&40&100&1000  \\
\noalign{\smallskip}\hline\noalign{\smallskip}
$P_{0/3}(x)$& 0.880&0.178&3.2$\times10^{-6}$&2$\times10^{-7}$&1.9$\times10^{-10}$\\
$\ep_{0/3}(x)$&-0.15 &$-$58.083&-99.72&-99.80&-99.86 \\
$P_{1/4}(x)$ &0.882&0.424&0.00108&0.000097&1.346 $\times 10^{-7}$\\
$\ep_{1/4}(x)$&-0.00033&-0.005&-2.93&-3.22&-0.4\\
$P_{2/5}(x)$&0.882&0.424&0.00162&0.00024&-9.318 $\times 10 ^{-7}$\\
$\ep_{2/5}(x)$&-0.00034&-0.024&45.1&141.8&-780\\
$exact$\cite{Bender_30}&0.8817&0.4240&0.001114&0.0001002&1.351275 $\times10^{-7}$\\
\noalign{\smallskip}\hline
\end{tabular}
\end{table}

\section{Nonlinear Schr\"{o}dinger equation}

The nonlinear Schr\"{o}dinger equation can be met in many chemical and physical problems.
Here we keep in mind the variant describing a spherically symmetric function $f = f(r)$,
with $r \geq 0$, satisfying the equation $E[f] = 0$ in the form
\be
\label{26}
 \frac{d^2f}{dr^2} + \frac{1}{r} \; \frac{df}{dr} \; - \; \frac{f}{r^2} + f -
f^3 = 0 \;  .
\ee
Under the boundary conditions
\be
\label{27}
  f(0) = 0 \; , \qquad f(\infty) = 1 \; ,
\ee
the solution to this equation represents a vortex line.

At small values of the dimensionless variable $r$, the solution is written
\cite{Yukalova_35,Yukalov_36} as the expansion in powers of $r^2$,
\be
\label{28}
f(r) \simeq cr \left ( 1 + a_1 r^2 + a_2 r^4 + a_3 r^6 + a_4 r^8 \right ) \qquad
( r \ra 0 ) \;   ,
\ee
where
$$
 a_1 = - \frac{1}{8} \; , \quad a_2 = \frac{1+8c^2}{192}  \; ,
\quad a_3 = -  \frac{1+80c^2}{9216} \; ,\quad
a_4 = \frac{1+656c^2+1152 c^4}{737280} \; ,
$$
and the value of $c$ is prescribed by the second of the boundary conditions (\ref{27}).
Here we take $c = 0.58319$, as defined by Ginzburg and Sobyanin \cite{Ginzburg_37}.

At large $r$, one has \cite{Yukalova_35} the asymptotic expansion
\be
\label{29}
f(r) \simeq 1 + \frac{b_1}{r^2}  + \frac{b_2}{r^4} + \frac{b_3}{r^6} \qquad
(r\ra \infty) \; ,
\ee
with
$$
 b_1 = -\; \frac{1}{2} \; , \qquad b_2 = -\; \frac{9}{8} \; , \qquad
b_3 = -\; \frac{161}{16} \;  .
$$

As an irrational factor, we can take the first-order root approximant
\be
\label{30}
f_{irr}^*(r) = \frac{cr}{\sqrt{1 + A r^2}} \;   ,
\ee
with $A = 0. 0163972$. Then, following the scheme of Sec. 2, we define the correcting
function $f(r)/ f_{irr}^*(r)$, expand it in powers of $r$, and construct the related
Pad\'{e} approximant
\be
\label{31}
  P_{2/2}(r) = \frac{1 + A_1 r^2 + A_2 r^4}{1 + B_1 r^2 + B_2 r^4} \; ,
\ee
in which
$$
A_1 = 0.0674195 \; , \qquad A_2 = 0.000899209 \; , 
$$
$$
B_1 = 0.184221 \; , \qquad B_2 = 0.00409531 \;   .
$$
Thus the corrected Pad\'{e} approximant is
\be
\label{32}
 f_4^*(r) = f_{irr}^*(r) P_{2/2}(r) \;  .
\ee

The accuracy of an approximate solution $f_{app}(r)$ to the considered equation, that
is denoted as
\be
\label{33}
  E[f(r)] = 0 \; ,
\ee
can be conveniently characterized by the {\it solution defect}
\be
\label{34}
  D[f_{app}(r) ] \equiv |\; E[f_{app}(r)] \; | \; ,
\ee
which defines the {\it maximal solution defect}
\be
\label{35}
 D[f_{app}] \equiv \sup_r D[f_{app}(r) ] \;  .
\ee

The found solution (\ref{32}) turns out to be very accurate, having the maximal defect
$D[f_4^*] = 0.0002$. The standard Pad\'{e} approximants are not applicable, since they
cannot satisfy the correct asymptotic behavior as $ r\rightarrow \infty$. Approximate
solutions, represented in the form of modified Pad\'{e} approximants \cite{Berloff_38},
are less accurate than the corrected approximant (\ref{32}). For example, the modified 
Pad\'{e} approximant
$$
 P'_{3/3}(r) = \left [
\frac{c_1 r^2 + c_2 r^4 + c_3 r^6}{1 + d_1 r^2 + d_2 r^4 + d_3 r^6}
\right ]^{1/2} \;  ,
$$
with
$$
c_1 = 0.340111\; , \qquad c_2 = 0.0745487 \; , \qquad c_3 = 0.0181768 \; ,
$$
$$
 d_1 = 0.469190 \; , \qquad d_2 = 0.0927255 \; , \qquad d_3 = 0.0181768 \;  ,
$$
has the maximal defect $D[P_{3/3}] = 0.04$, which is two orders larger than the maximal
solution defect of the approximant (\ref{32}).

It is possible to try Pad\'{e} approximants considered as fractional with respect to the powers
of expansions (\ref{28}) and (\ref{29}). For such fractional approximants we have
\begin{equation}
p_{2/2}(r)=\frac{0.58319 r \left(1+0.758279 r+0.600453 r^2+0.971322 r^3\right)}
{1+0.758279 r+0.725453 r^2+0.350178 r^3+0.566465 r^4}
\end{equation}
and
\begin{equation} 
\label{P}
p_{2/3}(r)= \frac{Q_2(r)}{Q_3(r)} \; ,
\ee
where
$$
Q_2(r) =
0.58319 r ( 1 + 0.691638 r + 0.307419 r^2 + 0.0996674 r^3 + 0.0234523 r^4 )
$$
$$
Q_3(r) =
1 + 0.691638 r + 0.432419 r^2 + 0.186122 r^3 + 0.058125 r^4 + 0.0136771 r^5 \; .
$$
The latter approximants can provide a reasonable accuracy for not too large $r$, although
they fail for $r \ra \infty$, not satisfying the boundary conditions (\ref{27}).

The results for different approximants are presented in Tables 4 and 5, from where it is 
seen that the corrected approximant $f^*_4(r)$ is the most accurate.

\begin{table}
\caption{Different types of Pad\'{e} approximants and the corrected approximant $f^*_4(r)$
for the nonlinear Schr\"{o}dinger equation, with the corresponding solution defects}
\label{Table 4}      
\centering
\begin{tabular}{lllllllll}
\hline\noalign{\smallskip}
$r$ & 0.1 & 1&2&3&5&7 \\
\noalign{\smallskip}\hline\noalign{\smallskip}
$p_{2/2}(r) $ & 0.0583 & 0.571&0.856&0.937&0.979&0.989 \\
$D[p_{2/2}(r)]$ & 0.049 & -0.12 & -0.097 & -0.021 & -0.00174 & -0.000303 \\
$p_{3/3}(r) $ & 0.0582 & 0.52 & 0.801 & 0.912 & 0.973 & 0.988 \\
$D[p_{3/3}(r)]$& -6.6$\times10^{-6}$ & -0.0067 & 0.0057 & 0.013 & 0.0058 & 0.0002\\
$P_{3/3}(r)$ & 0.0582 & 0.523 & 0.82 & 0.927 & 0.978 & 0.989\\
$D[P_{3/3}(r)]$ & 0.00016 & 0.031 & -0.044 & -0.017 & -0.00088 & -0.000078\\
$ f_4^*(r)$ & 0.0582 & 0.52 & 0.805 & 0.918 & 0.977 & 0.989\\
$D[f_4^*(r)]$ & 1.9$\times10^{-10}$ & 9.3$\times10^{-6}$ & 0.000032 & -0.000071 & -0.00020 & -0.00011 \\
\noalign{\smallskip}\hline
\end{tabular}
\end{table}

Several approximants, that are close to the numerical solution \cite{Berloff_38}, are shown 
in Fig.\ref{figd}: The Pad\'{e} approximant $p_{3/3}(r)$ (dot-dashed line), with the maximal 
error $0.8\%$; the Pad\'{e} approximant $P'_{3/3}(r)$ (dotted line), with the maximal 
error of $1.6\%$. The corrected  Pad\'{e} approximant is the most accurate, with the maximal 
error less than $0.4\%$.

\begin{figure}
\begin{center}
\includegraphics[width=14cm]{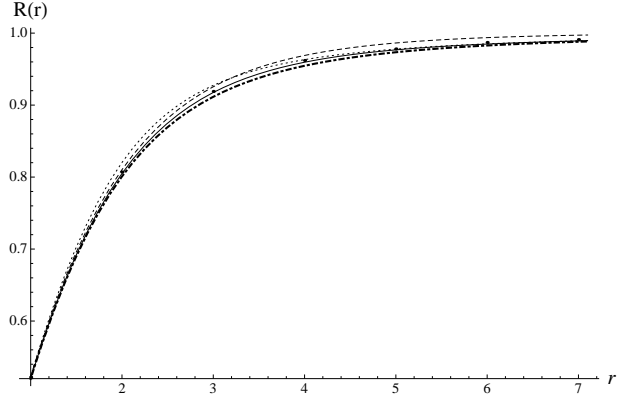}
\end{center}
\caption{
Different approximants are compared: The corrected Pad\'{e} approximant $f_4^*(r)$ (solid line);
the Pad\'{e} approximant $P'_{3/3}(r)$ (dotted line); the Pad\'{e} approximant $p_{3/3}(r)$
(dot-dashed line). Naive Berloff's formula 
$R(r)=\sqrt{\frac{r^2 \left(0.0286 r^2+0.3437\right)}{0.0286 r^4+0.3333 r^2+1}}$ from 
\cite{Berloff_38} is shown with dashed line. The numerical data from \cite{Berloff_38} are marked 
with dots: $R(0.1)=0.05825$, $R(1)=0.522$, $R(2)= 0.807$, $R(3)=0.919$, $R(4)= 0.962$, 
$R(5)=0.978$, $R(6)=0.987$, $R(7)= 0.991$.
}
\label{figd}
\end{figure}

\begin{table}[ht]
\caption{Different types of Pad\'{e} approximants and the corrected approximant $f^*_4(r)$
for the nonlinear Schr\"{o}dinger equation, with the corresponding errors.}
\label{Table 5}
\centering
\begin{tabular}{llllllll}
\hline\noalign{\smallskip}
$r$ &$p_{3/3}(r) $&$\ep_{p_{3,3}}$ & $P_{3/3}(r) $ & $\ep_{P_{3/3}}$ & $ f_4^*(r)$ &$\ep_{f_4^*(r)}$ & numerical  \\ \hline
0.1 & 0.0582 & -0.0065             & 0.0582 & -0.0064   &   0.0582 &-0.065  & 0.05825\\
1    & 0.52  & -0.46               & 0.523  & 0.26      &   0.52   & -0.37  & 0.522      \\
2    & 0.801 & -0.75               & 0.82   &  1.64     &   0.805  & -0.25  & 0.807 \\
3    & 0.912 & -0.81               & 0.927  &   0.92    & 0.918    & -0.05  & 0.919 \\
5    &0.973  & -0.47               & 0.978  & -0.022    & 0.977    &-0.13   & 0.978 \\
7    & 0.988 & -0.3                & 0.989  & -0.18     & 0.989    & -0.18  & 0.991 \\ 
\noalign{\smallskip}\hline
\end{tabular}
\end{table}

\section{Ruina-Dieterich equation}

The  Ruina-Dieterich  equation describes the law of  friction  between two solid surfaces
sliding against each other \cite{Scholz_39}. In dimensionless form the equation reads as
\be
\label{36}
 \frac{df}{dt} = b - f^{1-m} \;  ,
\ee
defining a semi-positive function $f = f(t)\geq 0$ of dimensionless time $t \geq 0$. Here
we consider the values $b = 0.526$ and $m = 3/2$. The initial condition is $f(0) = 0.5$.

In the asymptotic limit of short time,
\be
\label{37}
  f(t) \simeq f_{10}(t) \qquad ( t \ra 0) \; ,
\ee
the solution can be written as the expansion
\be
\label{38}
 f_{10}(t) = \sum_{n=0}^{10} a_n t^n \;  ,
\ee
with the coefficients
$$
 a_0 = \frac{1}{2} \; , \qquad a_1 = -0.888214 \; , \qquad a_2 = -0.628062 \; ,
$$
$$
a_3 = -0.853924 \; , \quad a_4 = -1.51297 \; , \quad a_5 = -3.06015 \; ,
\quad a_6 = -6.70249 \; ,
$$
$$
a_7 = -15.4836 \; , \quad a_8 = -37.1618 \; , \quad a_9 = -91.7923 \; ,
\quad a_{10} = -231.875 \;  .
$$
As is evident, the above series diverge.

Contrary to the previous two cases, the solution to this equation does not extend to
infinite times, but is limited from the right by a critical point $t_c = 0.329956$, where
\be
\label{39}
 f(t) \simeq f_{irr}(t) \qquad ( t \ra t_c - 0) \; ,
\ee
and in the vicinity of this point the function exhibits the irrational behavior in the form
of the root approximant
\be
\label{40}
  f_{irr}(t) = \frac{1}{2} \left ( 1 \; - \; \frac{t}{t_c} \right )^{2/3} \; .
\ee

Introducing the correcting function
\be
\label{41}
 C_{10}(t) = \frac{f_{10}(t)}{f_{irr}(t)} \;  ,
\ee
we expand the latter in powers of time, getting
\be
\label{42}
 C_{10}(t) \simeq \sum_{n=0}^{10} b_n t^n \qquad (t \ra 0 ) \;  ,
\ee
with the coefficients
$$
 b_0 = 1 \; , \qquad b_1 = 0.244047 \; , \qquad b_2 = 0.257547 \; ,
$$
$$
b_3 = 0.436291 \; , \qquad b_4 = 0.884281 \; , \qquad b_5 = 1.96929 \; ,
\qquad b_6 = 4.64878 \; ,
$$
$$
b_7 = 11.4182 \; , \qquad b_8 = 28.8636 \; , \qquad b_9 = 74.5732 \; ,
\qquad b_{10} = 196.0 \;  .
$$

On the basis of expansion (\ref{42}), we define the Pad\'{e} approximant
\be
\label{43}
 P_{5/5}(t) = \frac{\sum_{n=0}^5 c_n t^n}{\sum_{n=0}^5 d_n t^n} \;  ,
\ee
whose parameters are
$$
c_0 = 1 \; , \qquad c_1 = -6.89716 \; , \qquad c_2 = 16.4086 \; ,
$$
$$
c_3= -15.5576 \; , \qquad c_4 = 4.88054 \; , \qquad c_5 = -0.15572 \; ,
$$
$$
d_0 = 1 \; , \qquad d_1 = -7.14121 \; , \qquad d_2 = 17.8938 \; ,
$$
$$
d_3= -18.5216 \; , \qquad d_4 = 7.02358 \; , \qquad d_5 = -0.560998 \; .
$$

Thus the sought approximate solution is
\be
\label{44}
 f_{10}^*(t) = f_{irr}(t) P_{5/5}(t) \;  .
\ee
This solution very well approximates the sought function $f(t)$. The maximal deviation
of the corrected Pad\'{e} approximant (\ref{44}) from the exact numerical solution $f(t)$,
in the whole range of its definition, is
\be
\label{45}
 \sup_{t\in[0,t_c]} \; \left [ f(t) - f_{10}^*(t) \right ] = 0.0002 \; .
\ee

To study convergence, we consider the lower-order corrected approximants
\be
\label{444}
 f_{8}^*(t) = f_{irr}(t) P_{4/4}(t) \;  ,
\ee
with
$$
P_{4/4}(t)=\frac{1-5.37518 t+8.8021 t^2-4.37466 t^3+0.243023 t^4}
{1-5.61923 t+9.91591 t^2-5.78369 t^3+0.668044 t^4} \; ,
$$

\be
\label{4444}
 f_{6}^*(t) = f_{irr}(t) P_{3/3}(t) \;  ,
\ee
with
$$
P_{3/3}(t)=\frac{1-3.85376 t+3.51763 t^2-0.372822 t^3}
{1-4.09781 t+4.26014 t^2-0.793411 t^3} \; ,
$$
and
\be
\label{44444}
 f_{4}^*(t) = f_{irr}(t) P_{2/2}(t) \;  ,
\ee
with
$$
P_{2/2}(t)=\frac{1-2.33262 t+0.560183 t^2}{1-2.57667 t+0.931464 t^2} \; .
$$
The related results are presented in Table 6. These results demonstrate good convergence of 
the corrected approximants.

For comparison, we also study the Pad\'{e} approximants
$$
p_{2/2}(t)=\frac{0.5-1.99743 t+1.64601 t^2}{1-2.21843 t+0.60727 t^2} \; ,
$$
$$
p_{3/3}(t)=\frac{0.5-2.76077 t+4.42424 t^2-1.83935 t^3}{1-3.74512 t+3.45167 t^2-0.543545 t^3} \; ,
$$
$$
p_{4/4}(t)=\frac{0.5-3.52899 t+8.39113 t^2-7.53882 t^3+1.92114 t^4}
{1-5.28156 t+8.65607 t^2-4.62722 t^3+0.501267 t^4} \; ,
$$
and
$$
p_{5/5}(t)=\frac{0.5-4.30291 t+13.5796 t^2-19.0149 t^3+11.167 t^4-1.96629 t^5}
{1-6.82939 t+16.2834 t^2-15.9743 t^3+5.77314 t^4-0.478209 t^5}.
$$
The corresponding results are shown in Table 7. As is seen, the corrected approximants are 
more accurate than the Pad\'{e} approximants.

For further comparison, we also study the non-diagonal Pad\'{e} approximants
$$
p_{1/2}(t)= \frac{\frac{1}{2}-1.33466 t}{1-0.892888 t-0.330028 t^2}\; ,
$$
$$
p_{2/3}(t)=\frac{0.5-2.20204 t+2.19219 t^2}{1-2.62765 t+0.972664 t^2+0.135056 t^3} \; ,
$$
$$
p_{3/4}(t)=\frac{0.5-3.00676 t+5.5076 t^2-2.91787 t^3}{1-4.2371 t+4.74443 t^2-1.02208 t^3-0.0664454 t^4} \; ,
$$
and
$$
p_{4/5}(t)=\frac{0.5-3.79116 t+9.96764 t^2-10.4266 t^3+3.45104 t^4}{1-5.80588 t+10.8777 t^2-7.11483 t^3+1.03717 t^4+0.0348388 t^5}\,.
$$
The results are shown in Table 8.

\begin{table}
\caption{Self-similarly corrected Pad\'{e} approximants with their percentage errors}
\label{Table 6}     
\centering
\begin{tabular}{llllllllllll}
\hline\noalign{\smallskip}
$t$ &0.05&0.1&0.2&0.25&0.3&0.32\\
\noalign{\smallskip}\hline\noalign{\smallskip}
$f_4^*(t)$ & 0.454 & 0.404&0.286&0.21&0.114&0.0556 \\
$\ep_4^*(t)$ & 0.000016 &-$0.00015$&-0.02&-0.10&-0.71&-1.98 \\
$f_6^*(t)$&0.454&0.404&0.286&0.212&0.115&0.0562 \\
$\ep_6^*(t)$& 0.000020&0.000047 &$-$0.00072&-0.012&-0.2&-0.94\\
$f_8^*(t)$&0.454&0.404&0.286&0.212&0.115&0.0565\\
$\ep_8^*(t)$&0.000020&0.000049 &0.00014&$-$0.0010&$-$0.059&-0.46\\
$f_{10}^*(t)$&0.454&0.404&0.286&0.212&0.115&0.0566\\
$\ep_{10}^*(t)$&0.000020&0.000049&0.00018&0.00024&$-$0.016&$-$0.23\\
$exact$&0.453902&0.403853&0.286184&0.212307&0.114745&0.0567535\\
\noalign{\smallskip}\hline
\end{tabular}
\end{table}

\begin{table}
\caption{Diagonal Pad\'{e} approximants with their percentage errors.}
\label{Table 7}     
\centering
\begin{tabular}{llllllllllllll}
\hline\noalign{\smallskip}
$t$ &0.05&0.1&0.2&0.25&0.3&0.32\\
\noalign{\smallskip}\hline\noalign{\smallskip}
$p_{2/2}(t)$ & 0.454 & 0.404&0.287&0.214&0.126&0.0834 \\
$\ep_{2/2}(t) $&0.000023& 0.00108&0.12&0.88&9.55&46.9 \\
$p_{3/3}(t) $&0.454&0.404&0.286&0.213&0.118&0.068\\
$\ep_{3/3}(t)$&2.4$\times 10^{-6}$& 0.000013&0.0059&0.0972&2.53&19.73 \\
$p_{4/4}(t) $&0.454&0.404&0.286&0.212&0.116&0.0618\\
$\ep_{4/4}(t) $&2.4$\times 10^{-6}$&4.9 $\times 10^{-6}$&0.00033&0.011&0.70&8.92\\
$p_{5/5}(t) $&0.454&0.404&0.286&0.212&0.115&0.0591\\
$\ep_{5/5}(t) $&2.4$\times 10^{-6}$&4.8$\times 10^{-6}$&0.000046&0.0013&0.19&4.17 \\
\noalign{\smallskip}\hline
\end{tabular}
\end{table}

\begin{table}
\caption{Non-diagonal Pad\'{e} approximants with their percentage errors.}
\label{Table 8}      
\centering
\begin{tabular}{llllllllllllll}
\hline\noalign{\smallskip}
$t$ &0.05&0.1&0.2&0.25&0.3&0.32\\
\noalign{\smallskip}\hline\noalign{\smallskip}
$p_{1/2}(t)$ & 0.454 & 0.404&0.288&0.22&0.142&0.107 \\
$\ep_{1/2}(t) $&0.0009&0.02&0.76&3.61&23.6&88.8 \\
$p_{2/3}(t) $&0.454&0.404&0.286&0.213&0.121&0.0753\\
$\ep_{2/3}(t)$&3.61$\times 10^{-6}$&0.000138&0.034&0.357&5.55&32.74 \\
$p_{3/4}(t) $&0.454&0.404&0.286&0.212&0.116&0.0648\\
$\ep_{3/4}(t) $&2.382$\times 10^{-6}$&5.818 $\times 10^{-6}$&0.00168&0.0386&1.472&14.1\\
$p_{4/5}(t) $&0.454&0.404&0.286&0.212&0.115&0.0604\\
$\ep_{4/5}(t) $&2.381$\times 10^{-6}$&4.844$\times 10^{-6}$&0.000114&0.00436&0.407&6.47 \\
\noalign{\smallskip}\hline
\end{tabular}
\end{table}

Again, we see that the corrected Pad\'{e} approximants are essentially more accurate than the
standard Pad\'{e} approximants.

\section{Possible further extensions}

The described method of self-similarly corrected  Pad\'{e} approximants can be extended 
to other problems. Here we mention some of possible applications.

\subsection{Finding large-variable behavior}

The method can be used for finding the large-variable behavior of functions from their
small-variable asymptotic expansions. For instance, suppose we have a small-variable 
approximation $f_k(x)$ at $x \ra 0$. And we need to find out the behaviour of the function 
at large variables, when $x \ra \infty$. As is evident, the small-variable series $f_k(x)$ 
has no sense for $x \ra \infty$. But to find the large-variable exponent, it is possible to 
proceed as follows. We define the function
\be
\label{68}
  \bt_k(x) = \frac{d\ln f(x)}{d\ln x} 
\ee
expanding it in powers of $x$. Then, according to sec. 2, we derive the self-similarly 
corrected Pad\'e approximant
\be
\label{69}
  \bt_k^*(x) = \bt_{irr}^*(x) P_{N/N}(x) \;  .
\ee
Taking the limit
\be
\label{70}
 \bt_k = \lim_{x\ra\infty} \bt_k^*(x) \;  ,
\ee
we obtain the large-variable exponent defining the behaviour of the sought function 
at large variables as
\be
\label{71}
 f_k^*(x) \simeq B_k x^{\bt_k} \;  .
\ee

We have checked this way of defining the large-variable exponents for several physical 
problems and compared the accuracy of the corrected approximants with the standard 
Pad\'{e} approximants. As expected, the corrected approximants are always more accurate 
than the standard Pad\'{e} approximants and, moreover, exist in the cases where the 
standard Pad\'{e} approximants cannot be defined at all. Root approximants can also 
be used as an initial approximation for calculating critical exponents at phase transitions
\cite{Gluzman_42}. In order not to overload  the present paper, we do not go here into the 
details of calculating the large-variable and critical exponents, since our main aim has been 
to demonstrate the applicability of the method for an accurate solution of nonlinear 
differential equations.

\subsection{Partial differential equations}

As we have demonstrated, the method can be applied for solving nonlinear ordinary 
differential equations. Its generalization to partial differential equations can be done in 
the following cases. 

\vskip 2mm

(i) {\it Equation allows for the standard separation of variables}. Suppose we consider a 
partial differential equation for which the variables can be separated by the standard 
procedure \cite{Jeffreys_43}. Then the problem can be reduced to the set of equations in 
ordinary derivatives. For each of the separated equations, containing a single variable, 
it is straightforward to use the developed method. 

\vskip 2mm

(ii) {\it Functional separation of variables is allowed}. The reduction of a partial 
differential equation to several ordinary differential equations is also admissible under the
functional separation of variables \cite{Jia_43}. For example, we consider an equation for 
a function $f(x,t)$ of two variables. One says that the equation allows for the functional 
separation of variables if there exist functions $F(f)$, $\psi(x)$, and $\varphi(t)$, for which
$$
F(f) = \psi(x) + \varphi(t) \; .
$$
The separation is called additive for $F(f) = f$ and it is called product separation, if 
$F(f) = \ln f$. Then, instead of one equation in partial derivatives for $f$, one gets two 
equations in ordinary derivatives \cite{Jia_43}. 

More generally, when the sought function depends on several variables, say 
$f = f(x_1,x_2,\ldots)$, they can be functionally separated if there exists a function $F(f)$ 
such that
$$
F(f) = f_1(x_1) + f_2(x_2) + \ldots \; .
$$
Then one equation in partial derivatives separates in a set of equation in ordinary derivatives.

\vskip 2mm

(iii) {\it Generalized separation of variables is admissible}. For a function of two 
variables $f(x,t)$, this means the following. A partial differential equation can be reduced 
to two ordinary differential equations, when there exist functions $g(x/\varphi)$, $\varphi(t)$, 
and $y(t)$, such that
$$
f(x,t) = g(x/\varphi(t)) y(t) \; .
$$   
Again, instead of one equation in partial derivatives, one obtains two equations in ordinary 
derivatives \cite{Polyanin_44,Polyanin_45}.

\vskip 2mm

(iv) {\it Perturbation theory is used, starting from an approximation with separated variables}. 
Then the solution for the initial approximation can be constructed by employing the method of
corrected Pad\'{e} approximants for each of the separated equations.  

\vskip 2mm

(v) {\it Solving an eigenvalue problem, one is interested in eigenvalues}. Then it is
possible to use perturbation theory in powers of some parameter, considering the eigenvalue 
as a function of this parameter. For instance, we can be interested in the energy levels of
a three-dimensional Schr\"{o}dinger equation as a function $E(g)$ of a coupling parameter $g$.
In that case, the problem is reduced to studying the perturbative series for the eigenvalues
$E(g)$ depending on one variable $g$. Such perturbative series can be derived for linear as 
well as for nonlinear equations \cite{Yukalov_46,Courteille_47,Yukalov_48}.   

\vskip 2mm

(vi) {\it The main aim is the construction of an effective equation interpolating between 
small and large parameters}. The typical example is the construction of an effective 
Schr\"{o}dinger equation, with the energy term interpolating between weak and strong coupling 
\cite{Gautam_49}. In the latter case, one is looking for the expression of an effective 
energy interpolating between the weak-coupling Lee-Huang-Yang form 
\cite{Huang_50,Lee_51,Lee_52} and the strong-coupling limit corresponding to unitarity 
\cite{Adhikari_53,Ding_54}. The interpolation can be done by using a kind of a two-point 
Pad\'{e} approximation \cite{Gautam_49}. Because of the small number of the available 
interpolation terms, the used two-point Pad\'{e} approximation omits some of the coefficients 
of the two-point approximants. Also, the number of available terms is not sufficient for using 
corrected Pad\'{e} approximants. Nevertheless, despite the simplicity, the constructed 
interpolation for the energy describes reasonably well the crossover between weak-coupling and 
strong-coupling limits. Then this energy term is substituted into the equation that is solved 
numerically \cite{Gautam_49}. Although in this case, the method of self-similar approximants, 
because of the small number of the known asymptotic terms, cannot been used, this example 
illustrates how the method could be applied, provided a sufficient number of expansion terms 
would be available.

\section{Conclusion}

When a problem is expected to possess a solution with rational functional behavior,
this solution can be described well by Pad\'{e} approximants that provide the best
representation for rational functions \cite{Baker_1}. Although a solution exhibiting 
irrational functional behavior can also be approximated by the higher-order Pad\'{e} 
approximants, the results in reasonably low orders are not sufficiently accurate. 
We are presently focused upon improving the convergence rate of the standard
Pad\'{e} approximants for lower-orders as well as for higher orders. This problem is
of high importance in describing finite quantum systems \cite{Birman_40} and structured 
media \cite{Gluzman_41}.

An approach is presented allowing for the extension of the method of Pad\'{e}
approximants to the problems with irrational functional behaviour of solutions. The main
idea of the approach is in splitting the sought solution into two factors. One of
them is defined through self-similar root approximants taking into account irrational
functional behaviour. While the second factor is constructed as a Pad\'{e} approximant
corresponding to a rational function, or a function with reducible irrationality. This
method of corrected Pad\'{e} approximants is illustrated by finding approximate analytical
solutions to nonlinear differential equations typical of physics, and several branches of 
applied sciences. We present accurate solutions to Thomas-Fermi equation, 
nonlinear Schr\"{o}dinger equation, and Ruina-Dieterich equation. The method is shown to 
provide a very high accuracy of solutions, essentially better than that of the standard 
Pad\'{e} approximants. Since the formulation of the method is general, the described 
approach can be applied to any nonlinear ordinary differential equation. Possible use of
the method for partial differential equations is discussed.

\section*{Acknowledgements} 
One of the authors (V.I.Y.) acknowledges the help from E.P. Yukalova.

\vskip 1cm

{\parindent = 0pt
{\bf Conflict of interest}: The authors declare that they have no conflict of
interest. }


\vskip 2cm

\begin{thebibliography}{99}

\bibitem{He_1}
J.H. He,
{\it Int. J. Mod. Phys. B} {\bf 20}, 1141 (2006).

\bibitem{Baker_1}
G.A. Baker and P. Graves-Moris, 
{\it Pad\'{e} Approximants} (Cambridge University, Cambridge, 1996).

\bibitem{Epele_3}
L.N. Epele, H. Fanchiotti, C.A. Garcia Canal, and J.A. Ponciano,  
{\it Phys. Rev. A} {\bf 60}, 280 (1999).

\bibitem{Gluzman_7}
S. Gluzman and V.I. Yukalov,  
{\it Eur. J. Appl. Math.} {\bf 25}, 595 (2014).

\bibitem{Gluzman_8} 
S. Gluzman and V.I. Yukalov,  
{\it Mathematics} {\bf 3}, 510 (2015).

\bibitem{Yukalov_8}
V.I. Yukalov,
{\it Int. J. Mod. Phys. B} {\bf 3}, 1691 (1989).

\bibitem{Yukalov_11}
V.I. Yukalov,  
{\it Phys. Rev. A} {\bf 42}, 3324 (1990).

\bibitem{Yukalov_12}
V.I. Yukalov, 
{\it J. Math. Phys.} {\bf 32}, 1235 (1991).

\bibitem{Yukalov_13}
V.I. Yukalov,  
{\it J. Math. Phys.} {\bf 33}, 3994 (1992).

\bibitem{Yukalov_14}
V.I. Yukalov,
{\it Int. J. Mod. Phys. B} {\bf 7}, 1711 (1993).

\bibitem{Yukalov_15}
V.I. Yukalov and E.P. Yukalova,
{\it Int. J. Mod. Phys. B} {\bf 7}, 2367 (1993).

\bibitem{Struble_15}
R.A. Struble, ed.
{\it Nonlinear Differential Equations} (McGrow-Hill, New York, 1962).

\bibitem{Irwin_16}
M.C. Irwin,
{\it Smooth Dynamical Systems} (Academic, London, 1980).

\bibitem{Farmer_17}
J.D. Farmer,
{\it Physica D} {\bf 4}, 366 (1982). 

\bibitem{Peitgen_18}
H.O. Peitgen, ed. 
{\it Newton Method and Dynamical Systems} (Kluwer, Dordrecht, 1989).

\bibitem{Yukalov_19}
V.I. Yukalov, 
{\it Phys. Part. Nucl.} {\bf 50}, 141 (2019).

\bibitem{Yukalov_20}
V.I. Yukalov, E.P. Yukalova, and S. Gluzman, 
{\it Phys. Rev. A} {\bf 58}, 96 (1998).

\bibitem{Gluzman_21}
S. Gluzman and V.I. Yukalov,  
{\it Phys. Rev. E} {\bf 58}, 4197 (1998).

\bibitem{Yukalov_22}
V.I. Yukalov and S. Gluzman,  
{\it Physica A} {\bf 273}, 401 (1999).

\bibitem{Gluzman_23}
S. Gluzman and V.I. Yukalov,  
{\it J. Math. Chem.} {\bf 48}, 883 (2010).

\bibitem{Yukalov_24}
V.I. Yukalov and S. Gluzman,  
{\it Phys. Rev. D} {\bf 91}, 125023 (2015).

\bibitem{Gluzman_9}
S. Gluzman and V.I. Yukalov,  
{\it Eur. Phys. J. Plus} {\bf 131}, 340 (2016).

\bibitem{Yukalov_25}
V.I. Yukalov and E.P. Yukalova, 
{\it Chaos Solit. Fract.} {\bf 14}, 839 (2002).

\bibitem{Spruch_26}
L. Spruch,  
{\t Rev. Mod. Phys.} {\bf 63}, 151 (1991).

\bibitem{Fermi_27}
E. Fermi, 
{\it Z. Physik A} {\bf 48}, 73 (1928).

\bibitem{Baker_28}
E.B. Baker, 
{\it Phys. Rev.} {\bf 36}, 630 (1930).

\bibitem{Pindov_29}
G.I. Pindov and S.K. Pogrebnya,  
{\it J. Phys. B} {\bf 20}, 547 (1987).

\bibitem{Bender_30}
C.M. Bender, S.A. Orszag,  
{\it Advanced Mathematical Methods for Scientists and Engineers}
(McGraw-Hill, New York, 1978).

\bibitem{Sommerfeld_31}
A. Sommerfeld, 
{\it Rend. R. Accad. Lincei} {\bf 15}, 293 (1932).

\bibitem{Andrianov_32}
I.V. Andrianov and J. Awrejscewicz,  
{\it Phys. Lett. A} {\bf 319}, 53 (2003).

\bibitem{Desaix_32}
M. Desaix, D. Anderson, and M. Lisak,  
{\it Eur. J. Phys.} {\bf 25}, 699 (2004).

\bibitem{Bougoffa_33}
L. Bougoffa and R.C. Rach,  
{\it Rom. J. Phys.} {\bf 60}, 1032 (2015).

\bibitem{Sierra_34}
D. Sierra-Porta, M. Chirnos, and J. Stock,  
{\it Rev. Mexic. Fisica} {\bf 63}, 333 (2017).

\bibitem{Yukalova_35}
E.P. Yukalova, V.I. Yukalov, and S. Gluzman,  
{\it Ann. Phys. (N.Y.)} {\bf 323}, 3074 (2008).
 
\bibitem{Yukalov_36}
V.I. Yukalov, 
{\it Laser Phys.} {\bf 26}, 062001 (2016).
 
\bibitem{Ginzburg_37}
V.L. Ginzburg and A.A. Sobyanin,  
{\it J. Exp. Theor. Phys.} {\bf 56}, 455 (1982).

\bibitem{Berloff_38}
N.G. Berloff,  
{\it J. Phys. A} {\bf 37}, 1617 (2004).

\bibitem{Scholz_39}
C.H. Scholz,  
{\it Nature} {\bf 391}, 37 (1998).

\bibitem{Gluzman_42}
S. Gluzman and V.I. Yukalov,
{\it Eur. Phys. J. Plus} {\bf 132}, 535 (2017).

\bibitem{Jeffreys_43}
S.H. Jeffreys and B.S. Jeffreys, 
{\it Methods of Mathematical Physics} (Cambridge University, Cambridge, 1988). 

\bibitem{Jia_43}
H. Jia, W. Xu, and X. Zhao,
{\it J. Math. Anal. Appl.} {\bf 339}, 982 (2008).

\bibitem{Polyanin_44}
A.D. Polyanin, A.I. Zhurov, and A.V. Vyazmin,
{\it J. Non-Equil. Thermod.} {\bf 25}, 251 (2000).

\bibitem{Polyanin_45}
A.D. Polyanin, A.M. Kutepov, A.V. Vyazmin, and D.A. Kazenin,
{\it Hydrodynamics, Mass and Heat Traansfer in Chemical Engineering}
(Gordon and Breach, Singapore, 2000). 

\bibitem{Yukalov_46}
V.I. Yukalov and E.P. Yukalova,
{\it Laser Phys.} {\bf 5}, 154 (1995).

\bibitem{Courteille_47}
P.W. Courteille, V.S. Bagnato, and V.I. Yukalov,
{\it Laser Phys.} {\bf 11}, 659 (2001).

\bibitem{Yukalov_48}
V.I. Yukalov, E.P. Yukalova, and V.S. Bagnato,
{\it Phys. Rev. A} {\bf 66}, 043602 (2002). 

\bibitem{Gautam_49}
S. Gautam and S.K. Adhikari,
{\it Phys. Rev. A} {\bf 100}, 023626 (2019). 

\bibitem{Huang_50}
K. Huang and C.N. Yang, 
{\it Phys. Rev.} {\bf 105}, 767 (1957).

\bibitem{Lee_51}
T.D. Lee and C.N. Yang, 
{\it Phys. Rev.} {\bf 105}, 1119 (1957).

\bibitem{Lee_52}
T.D. Lee, K. Huang, and C.N. Yang, 
{\it Phys. Rev.} {\bf 106}, 1135 (1957).

\bibitem{Adhikari_53}
S.K. Adhikari and L. Salasnich, 
{\it Phys. Rev. A} {\bf 78}, 043616 (2008).

\bibitem{Ding_54}
Y. Ding and C. H. Greene, 
{\it Phys. Rev. A} {\bf 95}, 053602 (2017).

\bibitem{Birman_40}
J.L. Birman, R.G. Nazmitdinov, and V.I. Yukalov, 
{\it Phys. Rep.} {\bf 526}, 1 (2013).

\bibitem{Gluzman_41}
S. Gluzman, V. Mityushev, and W. Nawalaniec,  
{\it Computational Analysis of Structured Media} (Academic, London, 2017).


\end{thebibliography}
\end{document}